  \documentclass[preprint,11pt]{elsarticle}

 \journal{Physica D}

\usepackage{graphicx}
\usepackage{psfig}
\usepackage{color}

\begin{document}
 
 \begin{frontmatter}

\title{Separatrix crossing due to multiplicative colored noise.}
 
 \author{Jean-R\'egis Angilella }

\address {Normandie Universit\'e, UNICAEN, UNIROUEN, ABTE, ESIX Cherbourg, Caen 14000, France.}

\vskip.5cm
\begin{abstract}
{ 
 The effect of weak multiplicative colored noise on the dynamics of a Hamiltonian system is studied by means of asymptotic methods, in the vicinity of homoclinic or heteroclinic trajectories. A general expression for the probability of noise-induced separatrix crossing   is obtained, and is illustrated by means of a two-well Duffing oscillator. It shows how weak noise can significantly affect the dynamics near separatrices. In addition,  the influence of the degree of non-linearity  of the noise amplitude is discussed.
 
}
\end{abstract}
 
\maketitle
   



 
  \end{frontmatter}

\section{Introduction}
  
  Noise, even weak, can have tremendous effects on dynamical systems. This is a well known phenomenon which has been studied long ago, and in many contexts  \cite{Moss1989book,Landa1996Pendulum,Landa1997Pendulum,Landa1997SimpleSystems,Mallick2004,Liu2004,  Zhu2005,Ibrahim2006}. Physical systems are particularly sensitive to noise in the vicinity of unstable equilibrium positions. Here, the dominant forces almost balance each other and noise can thrust the system towards unpredictable directions. In the presence of homoclinic (or heteroclinic) cycles relating such equilibrium points, weak noise can create homoclinic bifurcations leading to an extremely complex dynamics. The occurrence of this complexity can be predicted by means of stochastic Melnikov functions \cite{Bulsara1990,Schieve1990,Frey1993,Franaszek1996,Franaszek1998,Simiu1996Melnikov}, which emerge as soon as one calculates the jump of undisturbed Hamiltonian for a particle moving near the separatrix. Such Melnikov functions have been  used in the past to derive necessary conditions for noise-induced escapes \cite{Simiu1996Melnikov,Simiu1996}. However, in spite of these active researches, the probabilistic dynamics in Hamiltonian system submitted to noise is still poorly understood, especially in the case of colored noise. 

In many situations of interest, noise takes the form of a piecewise regular forcing, which changes randomly and at random times.  These time intervals being finite, noise has a non-zero autocorrelation length and enters into the category of colored noise. 
  In a recent paper \cite{Angilella2019PRE}, an expression for the probability of noise-induced separatrix crossing has been derived for any two-dimensional Hamiltonian system submitted to  an additive noise of this kind  (Kubo-Anderson process).  The aim of the present paper is to generalize this result to multiplicative noise. Indeed, many physical systems are submitted to such noise, which is by essence non-uniform in the phase space. Random terms being proportional to some function of the state variables, some portions of the phase space might be free of noise, whereas other portions are noisy. The transition from the former to the latter can be abrupt if the amplitude of the noise is a non-linear function of the state variables.  For example, a pendulum attached to a non-inertial reference frame undergoing piecewise-constant accelerations will enter into this category. In another context, particles suspended in a turbulent flow modelled by means of a stochastic model \cite{Kallio1989,Graham1996}, are also submitted to a random force of this kind.

We have therefore investigated the effect of weak,  colored and multiplicative  noise on the dynamics of a system  in the vicinity of homoclinic or heteroclinic trajectories. To be precise, we consider  a Hamiltonian system with two degrees-of-freedom $\mathbf{r} = (q,p)$ and undisturbed Hamiltonian $H(\mathbf{r})$. In the absence of perturbation (either deterministic or random), trajectories in the phase space correspond to $H(\mathbf r(t)) = constant$.  When perturbations are present, the Hamiltonian is no longer conserved along trajectories
  $\mathbf r(t)$, but one can still use $H(\mathbf r(t))$ to determine in which portion of the phase space  the state-variable $\mathbf r(t)$ is located.
The calculation of the variations of $H$  along a {perturbed} trajectory $\mathbf r(t)$   therefore allowed us to quantify the effect of noise on the dynamical system.   Section \ref{theogene} presents this methodology, as well as the  calculations and the main results. It is followed by a section devoted to the two-well Duffing equation, where theoretical results are compared to numerical simulations. It also contains a discussion about the effect of the non-linearity of the noise amplitude. Conclusions and perspectives are drawn in section \ref{conclus}.

  \section{General considerations}
  \label{theogene}

We consider a perturbed Hamiltonian system of the form
\begin{eqnarray}
\label{qdot}
\dot q &=& \frac{\partial H}{\partial p} + \Lambda \, U_1(\mathbf r) + \varepsilon_1 f_1(\mathbf r) \, \xi_1(t),\\
\dot p &=& -\frac{\partial H}{\partial q} + \Lambda \, U_2(\mathbf r) + \varepsilon_2 f_2(\mathbf r) \, \xi_2(t),
\label{pdot}
\end{eqnarray}
  where $\mathbf r(t) = (q(t),p(t))$ is the state variable, $H(\mathbf r)$ is the Hamiltonian, $(U_1,U_2)$ and $(f_1,f_2)$ are deterministic vector fields and $\varepsilon_i$ and $\Lambda$ are positive constants. Terms   $\Lambda \, U_i$ can be thought of as a deterministic perturbation of the Hamiltonian system (e.g. friction, gravity, etc.), whereas 
  $\varepsilon_i f_i(\mathbf r) \, \xi_i(t)$ are random perturbations.   The functions $\xi_i(t)$ are assumed to be random processes, taking constant random values $\xi_{ik}$ during time intervals $[\tau_k,\tau_{k+1}]$, where $\tau_k$ are random times. We assume that the time intervals $\delta\tau_k = \tau_{k+1}-\tau_k$ have an exponential distribution, to manifest the fact that noise is memoryless. 
Ensemble averages will be denoted by $\langle \cdot \rangle$.   
  The average time duration $\langle \delta\tau_k \rangle$ is denoted $\delta\tau$ in the following. Random processes $\xi_i(t)$ are therefore colored Kubo-Anderson noises   \cite{Kubo1954,Anderson1954,Brissaud1974}, with an exponential auto-correlation function. In addition, we assume that the random pulses are centred and normalized, that is  $\langle \xi_{ik} \rangle =0$ and $\langle \xi_{ik}^2 \rangle = 1$.
  
  \begin{figure}
 \centerline{\includegraphics[scale=0.8]{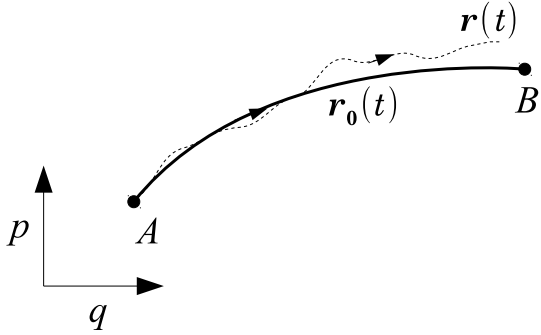}}
  \caption{ Sketch of a heteroclinic trajectory (separatrix) between two hyperbolic saddle points $A$ and $B$, in the unperturbed Hamiltonian system. Vector $\mathbf r_0(t)$ corresponds to a trajectory along the separatrix, and $\mathbf r(t)$ is a solution of the perturbed system. }
  \label{SketchSep}
  \end{figure}

  In the absence of either deterministic or stochastic perturbations ($\Lambda=\varepsilon_i = 0$), trajectories in the phase space correspond to $H(\mathbf r(t)) = constant$. When perturbations are present,  we use $H(\mathbf r(t))$ to determine the location of the state-variable in the phase space. If $AB$ denotes  a heteroclinic trajectory of the undisturbed system (Fig. \ref{SketchSep}), then $H(\mathbf r(t)) < H(A)$ means that $\mathbf r(t)$ is on the left-hand size of arc $AB$, and vice versa. Therefore, by examining the jump of Hamiltonian
  \begin{equation}
  \Delta H = H(\mathbf r(\tau_b)) - H(\mathbf r(\tau_a))
  \label{DeltaHdef}
  \end{equation}
  between two arbitrary times $\tau_a$ and $\tau_b$, one can check whether the state variable $\mathbf r(t)$ crossed the separatrix during the time interval $[\tau_a,\tau_b]$ (see for example Refs.\ \cite{Guckenheimer1983,Kuznetsov1997}). Here, the times $\tau_a$ and $\tau_b$ will be defined as the times where $\mathbf r(t)$ passes nearest to $A$ and $B$ respectively. Injecting Eqs.\ (\ref{qdot}) and (\ref{pdot}) into (\ref{DeltaH}) leads to \cite{Angilella2019PRE}
  \begin{equation}
  \Delta H = \Delta H^0 + \varepsilon_1 \int_{\tau_a}^{\tau_b} H_{,q} f_1(\mathbf r(t))
   \, \xi_1(t) \, dt
   + \varepsilon_2 \int_{\tau_a}^{\tau_b} H_{,p} f_2(\mathbf r(t))
   \, \xi_2(t) \, dt
\label{DeltaH}
\end{equation}   
  where commas indicate partial derivations and 
  \begin{equation}
  \Delta H^0 = \Lambda \int_{\tau_a}^{\tau_b} \nabla H \cdot \mathbf U \, dt
\label{DeltaH0}
  \end{equation}
  is the jump of Hamiltonian under the effect of the sole deterministic perturbation, which can be approximated by:
   \begin{equation}
  \Delta H^0 \simeq \Lambda \left( \int_{A}^{B} U_2(q,p) \, dq -  \int_{A}^{B} U_1(q,p) \, dp \right)
\label{DeltaH0bis}
  \end{equation}
  where integrations are done over the arc $AB$.
  
  \section{Calculation of the  jump of Hamiltonian}
  
\subsection{Approximations near the separatrix}

     To exploit further Eq. (\ref{DeltaH}), we need to calculate the two time integrals. Because the trajectory considered here is close to separatrix $AB$, as both $\Lambda$ and 
$\varepsilon_i$'s are small, we write  \cite{Guckenheimer1983}:
$$
\mathbf r(t) \simeq \mathbf r_0(t-t_0)
$$
where $\mathbf r_0(t)=(q_0(t),p_0(t))$ is a solution of the undisturbed system, that is $\dot q_0 = H_{,p}$ and      $\dot p_0 = -H_{,q}$ with $\mathbf r_0(-\infty) = A$ and  $\mathbf r_0(+\infty) = B$. In addition, we introduced the time $t_0$ where $\mathbf r(t_0)$ is nearest to $\mathbf r_0(0) \in ]A,B[$. This leads to:
  \begin{equation}
  \Delta H = \Delta H^0 - \varepsilon_1 \int_{\tau_a}^{\tau_b} \dot p_0(t-t_0) f_1(\mathbf r_0(t-t_0))
   \, \xi_1(t) \, dt
   + \varepsilon_2 \int_{\tau_a}^{\tau_b} \dot q_0(t-t_0) f_2(\mathbf r_0(t-t_0))
   \, \xi_2(t) \, dt.
\label{DeltaH2}
\end{equation}   
Noise being piecewise constant, the first integral can be split into a sum of elementary integrals of  $\dot p_0 f_1$ over $[\tau_k,\tau_{k+1}]$. In addition, if $\dot p_0$ keeps a constant sign on $[\tau_k,\tau_{k+1}]$, these elementary integrals can be calculated by the theorem of the mean value:   
$$
 \int_{\tau_k}^{\tau_{k+1}} \dot p_0(t-t_0) f_1(\mathbf r_0(t-t_0))
    \, dt = f_1( \mathbf r_0(\tau_{k+1/2}-t_0) )  \int_{\tau_k}^{\tau_{k+1}} \dot p_0(t-t_0)  
    \, dt  =  f_1( \mathbf r_0(\tau_{k+1/2}-t_0) ) \, \delta p_{0k},
$$
where $\tau_{k+1/2} \in ]\tau_k,\tau_{k+1}[$ and $\delta p_{0k} = p_0(\tau_{k+1}-t_0) -  p_0(\tau_{k}-t_0)$. The same treatment being done for the second integral, we get: 
    \begin{equation}
  \Delta H = \Delta H^0 - \varepsilon_1 \sum_{k=0}^{N-1} \xi_{1k}    f_1( \mathbf r_0(\tau_{k+1/2}-t_0) ) \, \delta p_{0k}
   + \varepsilon_2 \sum_{k=0}^{N-1} \xi_{2k}    f_2( \mathbf r_0(\tau_{k+1/2}-t_0) ) \, \delta q_{0k}.
\label{DeltaH3}
\end{equation}
We introduce the unit vector perpendicular to the separatrix at point $\mathbf r_0(\tau_k-t_0)$, approximated as:
$$
 \mathbf n_k =  (n_{1k},n_{2k}) =\Big(-\delta p_{0k},\delta q_{0k} \Big)/\delta s_k
$$
where $\delta s_k$ is the discrete arc-length element: $ \delta s_k^2 = \delta p_{0k}^2 + \delta q_{0k}^2$.
The jump of Hamiltonian therefore takes the form of a sum of random variables $X_k$:
  \begin{equation}
  \Delta H = \Delta H^0 +  \sum_{k=0}^{N-1}  X_k
\label{DeltaH4}
\end{equation}
where
 \begin{equation}
X_k = \sum_{i=1,2} \varepsilon_i  \xi_{ik}  \, n_{ik} \, f_{ik} \, \delta s_{k} 
 \end{equation}
and $f_{ik} =  f_i( \mathbf r_0(\tau_{k+1/2}-t_0) ) $. Being a sum of a large number of random variables (apart from the deterministic term $\Delta H^0$), the jump of Hamiltonian $\Delta H$ will be calculated below by means of the generalized central limit theorem.  

The times $\tau_k$ are decorrelated from the amplitudes $\xi_{ik}$ of the random force. Hence, the average of $X_k$ is:
$$
\langle X_k \rangle =  \sum_{i=1,2} \varepsilon_i \langle \xi_{ik} \rangle \, \langle  \, n_{ik} \, f_{ik} \, \delta s_{k} \rangle = 0
$$ 
since $ \langle \xi_{ik} \rangle = 0$. Following the methodology of Ref.\ \cite{Angilella2019PRE} we introduce:
\begin{equation}
Z_N = \frac{1}{S_N}  \sum_{k=0}^{N-1}  X_k 
\end{equation}
where $S_N^2 =  \sum_{k} \langle X_k^2 \rangle$. If $X_k$ satisfies the Lyapunov condition, i.e. if there exists $d > 0$ such that
\begin{equation}
\frac{1}{S_N^{2+d}}\sum_{k=0}^{N-1} \langle| X_k|^{2+d} \rangle \to  0
\label{lyapu}
\end{equation} 
as $N \to \infty$, then
 the random variable $Z_N$ converges to a centered Gaussian variable $Z$ with unit variance as $N \to \infty$. 
This allows us to approximate the jump of Hamiltonian as a Gaussian variable:
\begin{equation}
  \Delta H \simeq \Delta H^0 + \sigma \, Z,
\end{equation}
where $\sigma$ is the limit of $S_N$ as $N \to \infty$.

\subsection{Calculation of the variance of $\Delta H$}

The sum of the variances $S_N^2$ can be derived in a compact form in the case where the times of discontinuities $\tau_k$, their increments $\delta\tau_k$, and the amplitudes of the random forcing term $\xi_{ik}$ are independent. We follow the same main lines as in Ref. \cite{Angilella2019PRE}. First, by noticing that $\langle \xi_{1k} \xi_{2k} \rangle = 0$, and using  $\langle \xi_{ik}^2 \rangle = 1$, we get:
 \begin{equation}
\langle X_k^2 \rangle = \sum_{i=1,2} \varepsilon_i^2 \langle   n_{ik}^2 \, f_{ik}^2 \, \delta s_{k} ^2 \rangle.
 \end{equation}
By definition, the discrete arc-length element $\delta s_k$ is related to the velocity modulus on the separatrix by
$$
\delta s_k = u_{0k} \, \delta\tau_k
$$
where $u_{0k}=|\mathbf {\dot r}_0(\tau_k-t_0)|$. Also, the normal vector components $n_{ik}$, as well as $f_{ik}$ and $u_{0k}$ are functions of the times $\tau_k$, and are  decorrelated from $\delta\tau_k$. Therefore,
  $$
\langle   n_{ik}^2 \, f_{ik}^2 \, u_{0k}^2 \delta \tau_{k} ^2 \rangle 
= 2  \langle   n_{ik}^2 \, f_{ik}^2 \, u_{0k}^2  \rangle  \delta \tau^2
$$
where $\delta\tau =  \langle \delta \tau_{k}  \rangle$ and we made use of $\langle \delta \tau_{k} ^2 \rangle = 2 \delta\tau^2$ for exponentially distributed time increments. Finally, we have
\begin{equation}
S_N^2 =  \sum_{k} \langle X_k^2 \rangle = 2 \delta \tau \sum_i \sum_k  \varepsilon_i^2 \langle   n_{ik}^2 \, f_{ik}^2 \, u_{0k}^2  \rangle  \delta \tau .
\end{equation}
The sum over $k \in [0,N-1]$ in this last equation is of the form:
\begin{equation}
\langle \sum_k  g(\tau_k)   \delta \tau \rangle =  \langle \sum_k  g(\tau_k)   \delta \tau_k \rangle +  \langle \sum_k  g(\tau_k)   (\delta\tau - \delta \tau_k) \rangle.
\label{gtauk}
\end{equation}
The first sum on the right-hand-side of (\ref{gtauk}) is a Riemann sum that converges to $\int_{\tau_a}^{\tau_b} g(t) dt$ if $g(t)$ is integrable. The second sum is zero because $\tau_k$ is decorrelated from $\delta\tau_k$. We therefore conclude that, as $N \gg 1$, $S_N^2$ can be approximated by
\begin{equation}
\sigma^2 \simeq 2 \delta \tau \sum_{i=1,2}   \varepsilon_i^2  \int_{\tau_a}^{\tau_b} n_i^2(\mathbf r_0(t-t_0))
\, f_i^2(\mathbf r_0(t-t_0)) \, |\mathbf {\dot r}_0(\tau_k-t_0)|^2 \, dt.
\end{equation}
Alternatively, introducing the infinitesimal arc-length $ds$ along arc $AB$, defined by $ds = |\mathbf {\dot r}_0(\tau_k-t_0)| dt$:
\begin{equation}
\sigma^2 \simeq  2 \delta \tau \sum_{i=1,2}   \varepsilon_i^2  \int_{A}^{B} n_i^2(s)
\, f_i^2(s) \, |\nabla H(s)| \, ds. 
\label{sig2a}
\end{equation}
This expression is of interest since it involves the normals to the separatrix, which can lead to simple expressions for straight or circular separatrices, or in the case of isotropic noise. A useful alternative expression can be obtained in terms of integrals of $\dot p \, dp$ and $\dot q \, dq$, by noticing that $n_1  |\nabla H|=-\dot p$ and  $n_2  |\nabla H|=\dot q$, so that:
\begin{equation}
\sigma^2 \simeq  2 \delta \tau \Big(   \varepsilon_1^2  \int_{A}^{B}   f_1^2(q,p) \, \dot p \, dp +  \varepsilon_2^2  \int_{A}^{B}   f_2^2(q,p) \, \dot q \, dq
 \Big).
\label{sig2}
\end{equation}

  
   \subsection{Probability of noise-induced crossing}

Finally, having shown that the jump of Hamiltonian $\Delta H$ is a random variable that can be approximated by a Gaussian variable with mean $\Delta H^0$ (Eqs.\ (\ref{DeltaH0}) or (\ref{DeltaH0bis})) and variance $\sigma^2$ (Eq.\ (\ref{sig2})), we can determine the probability of noise-induced separatrix crossing in the same manner as for the case of additive noise \cite{Angilella2019PRE}. Noise-induce crossing corresponds to $\Delta H$ and $\Delta H^0$ having opposite signs, i.e. the effect of noise is opposed to the effect of the deterministic  perturbation. The probability of this event is:
\begin{equation}
 P =  \frac{1}{2} \mbox{erfc}\left( \frac{|\Delta H^0|}{\sigma \sqrt 2} \right).
\label{pb1}
\end{equation}
Eq.\     (\ref{pb1}), together with (\ref{sig2}) (or, alternatively (\ref{sig2a})), is the main result of this work, and will be illustrated in the following section. It differs from the additive case \cite{Angilella2019PRE}  in that $f_1$ and $f_2$ are now averaged in the calculation of the variance (\ref{sig2}).
Note that, to obtain this result, we have assumed that   $\dot p_0$ and $\dot q_0$  kept a constant sign on $[\tau_k,\tau_{k+1}]$. This assumption is needed to apply the theorem of the mean value to  elementary integrals of $ \dot p_0 f_1$ and $\dot q f_2$ over $[\tau_k,\tau_{k+1}]$ (see Eq.\ (\ref{DeltaH2}) and below). This hypothesis is not so restrictive, since the intervals of integration are small. In practice, it is satisfied in almost all intervals $[\tau_k,\tau_{k+1}]$.

 \section{Application to the Duffing equation} 
\label{secDuffing}
 
\subsection{Multiplicative noise with a linear amplitude}

\begin{figure}
 \centerline{\includegraphics[scale=0.7]{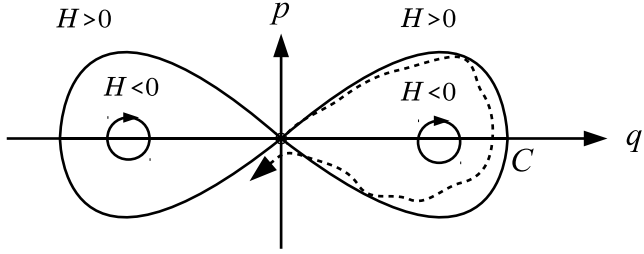}}
  \caption{Sketch of the phase portrait of the unperturbed inverted pendulum. The dashed line is a realization of the perturbed system.   }
  \label{SketchDuffing}
  \end{figure}

The probability calculated above can be applied to a wide variety of dynamical systems. Here we chose to illustrate these results by means of an inverted pendulum, which, in the limit of weak amplitudes, can be modelled as a two-well Duffing equation. It has been shown that additive and multiplicative noises in this family of dynamical systems had rather different effects \cite{Xu2011}.  The pendulum considered here is assumed to be composed of a rigid stick with length $l$, on the top of which a tiny object with mass $m$ has been fixed. It rotates around a fixed horizontal axis coinciding with its lower extremity, and a spring with stiffness $k$ exerts a torque $-k \theta$ that forces the pendulum back to its vertical position, where $\theta(t)$ denotes the angle between the vertical axis and the stick. In addition,  viscous friction with coefficient $b$ is present, together with a multiplicative random torque proportional to the angle $\theta$ (noise with a "linear amplitude"). The corresponding dynamical equation is
\begin{equation}
m l^2 \ddot \theta(t) = - k \, \theta(t) + m g l \sin \theta(t) - b \, \dot \theta(t) + \varepsilon \, \theta(t) \, \xi(t),
\label{InvertedPendulum}
\end{equation}
where $\varepsilon$ is a positive constant.
Assuming that $|\theta|$ is small we make use of the approximation $\sin\theta \simeq \theta - \theta^3/6$ (in practice this approximation is acceptable as long as $|\theta| <  \pi/4$). The dynamics of the pendulum therefore reads (removing time dependence for simplicity)
\begin{equation}
m l^2 \ddot \theta =  (mgl-k) \, \theta - \frac{1}{6} m g l \,  \theta^3 - b \, \dot \theta + \varepsilon \, \theta \, \xi.
\end{equation}
It can be re-written as:
\begin{equation}
\ddot \theta  =  \alpha \, \theta - \beta \, \theta^3 - \Lambda \, \dot \theta + \varepsilon_2 \, \theta \, \xi
\label{Duffing1}
\end{equation}
where $\alpha = (mgl-k)/(m l^2)$, $\beta = g/(6 l)$, $\Lambda = b/(m l^2)$ and $\varepsilon_2 = \varepsilon/(m l^2)$. This is a perturbed hamiltonian system of the form (\ref{qdot})-(\ref{pdot}) with $(q,p)=(\theta,\dot\theta)$ and
\begin{equation}
H(q,p) = -\frac{1}{2} \alpha \, q^2 + \frac{1}{4} \beta \, q^4 + \frac{1}{2}  \, p^2
\label{DuffingHamilt}
\end{equation}
and with a deterministic perturbation corresponding to $U_1(q,p) = 0$ and $U_2(q,p) = -p$. 
In this work, we will assume that $k < m g l$, so that $\alpha > 0$.
The phase portrait of the unperturbed 
system is obtained from the lines $H = constant$, and is sketched in Fig.  \ref{SketchDuffing}: it corresponds to a pair of homoclinic trajectories attached to a  hyperbolic saddle point located at the origin, and forming loops around elliptic points located at $q=\pm \sqrt{\alpha/\beta}$ and $p=0$. Separatrices intersect the $q$ axis at $q=\pm \sqrt{2\alpha/\beta}$.
When friction is present, the noise-free dynamics corresponds to trajectories spiraling towards the elliptic points, which are asymptotically stable
in this case. Any trajectory located inside one of the homoclinic loops will remain in this loop for all times.
 Noise, even weak, might break this picture. The probability that a pendulum, released near the origin and inside a homoclinic loop, exits this loop can be non-zero. Such a trajectory is sketched in Fig.  \ref{SketchDuffing}. According to the theory presented above, the calculation of this probability is given by Eq.\     (\ref{pb1}). The jump of Hamiltonian in the absence of noise reads (from Eq. (\ref{DeltaH0bis})):
$$
\Delta H^0 = - \Lambda \int_A^B p \, dq =  - 2 \Lambda \int_A^C p \, dq
$$
for symmetry reasons, where $C=(\sqrt{2\alpha/\beta},0)$ is the intersection between the $q$ axis and the homoclinic loop $AB$ (with $B=A$ here). On eliminating $p$ since $H(p,q)=0$ along $AC$ we get 
\begin{equation}
\Delta H^0 = - \frac{4}{3} \Lambda \, \frac{\alpha^{3/2}}{\beta}.
\label{DeltaH0Duffing}
\end{equation}
The sign of the deterministic jump of Hamiltonian indicates that friction drives the pendulum towards the interior of the homoclinic loops, as expected.
One can check that the Lyapunov condition is satisfied here (see below). The jump of Hamiltonian can therefore be approximated by a Gaussian variable, with standard deviation $\sigma$ given by Eq.\  (\ref{sig2}). Again, using $H(q,p)=0$ to eliminate $p$, we get:
\begin{equation}
\sigma = \frac{4\sqrt{2}}{\sqrt {15}} \,  \sqrt{\delta \tau} \, \varepsilon_2 \, \frac{\alpha^{5/4}}{\beta}.
\label{sigDuffing}
\end{equation}
Finally, by injecting Eqs.\ (\ref{DeltaH0Duffing}) and (\ref{sigDuffing}) into (\ref{pb1}) we obtain the probability of  noise-induced separatrix crossing for the inverted pendulum:
\begin{equation}
P = \frac{1}{2} \mbox{erfc}\left( \frac{\sqrt{15}}{6}\,  \frac{ \Lambda \,\alpha^{1/4}}{\varepsilon_2 \, \sqrt{\delta \tau}}   \right).
\label{pbDuffing}
\end{equation}
Note that it is independent of the coefficient $\beta$ of the cubic term in the dynamical equation (\ref{DuffingHamilt}). 
   \begin{figure}
 \centerline{\includegraphics[scale=0.55]{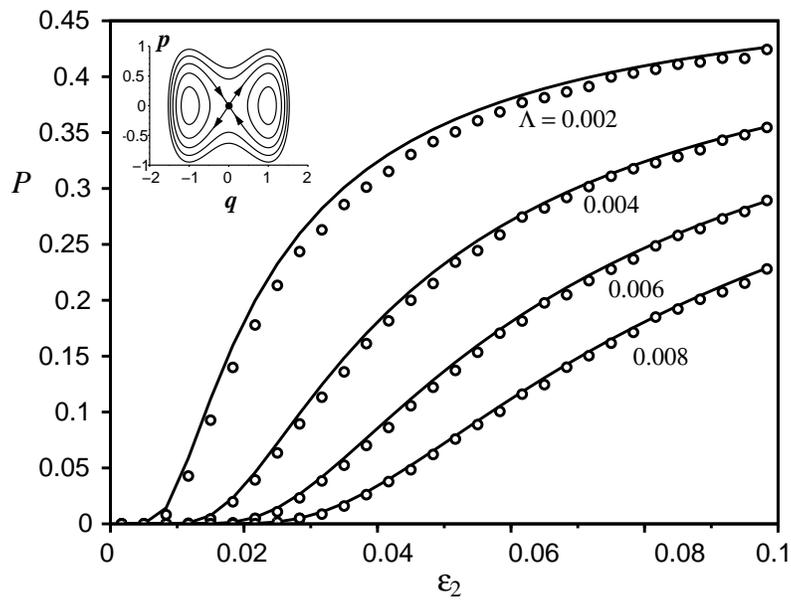}}
  \caption{Probability of noise-induced separatrix crossing for the inverted pendulum (Eq. (\ref{InvertedPendulum})) (multiplicative noise with linear amplitude). Solid line: theoretical result (\ref{pbDuffing}). Circles:  probability obtained from numerical solutions of Eq.\  (\ref{Duffing1}) involving 50000 stochastic trajectories.   }
  \label{ProbaDuffing}
  \end{figure}
 This theoretical probability is plotted versus $\varepsilon_2$ in Fig.\ \ref{ProbaDuffing}, in the case $\alpha=\beta=1\, s^{-2}$, $\delta\tau=0.01\,s$ and for various values of the friction coefficient $\Lambda$ (in units of $s^{-1}$). In addition, results from numerical solutions of Eq.\  (\ref{Duffing1}) involving 50000 stochastic trajectories starting at point $(0.001,0.001)$, are shown (circles). This numerical probability is the percentage of trajectories that join the zone $q < 0$ right after performing their first loop near the right-hand side homoclinic orbit. We observe that the agreement between this percentage and the analytical probability is correct.

\subsection{Multiplicative noise with a non-linear amplitude}

The theory presented above allows to study the effect of non-linear multiplicative noises, i.e. random forcing where either $f_1$ or $f_2$ is a non-linear function of $\mathbf r$.  
In the case of the Duffing oscillator studied in the previous section, the random force is weaker near the origin than near point $C$, and this non-uniformity can be even more pronounced if the amplitude of this force is a non-linear function of $(q,p)$: this might affect the probability of noise-induced crossing. To quantify this effect we have replaced the term $\varepsilon_2 \theta \xi$ in Eq.\ (\ref{Duffing1}) by $\varepsilon_2 \theta^\gamma \xi$, where $\gamma$ is a real number. This corresponds to $f_2(\mathbf r) = q^\gamma$.

Prior using the general expression of the probability calculated in section \ref{theogene}, we must check that the Lyapunov condition (\ref{lyapu}) is satisfied. To this end, we first note that $S_N \simeq K \sqrt {\delta \tau}$, where $K$ is  independent of $N$. This is a consequence of Eq.\  (\ref{sig2}). Also, $\varepsilon_1=0$ for the Duffing equation, so that:
\begin{equation}
|X_k|^{2+d} = \varepsilon_2^{2+d} \, | \xi_{2k}|^{2+d} \, |n_{2k} \, f_{2k} \, u_{0k} |^{2+d} \, |\delta\tau_k |^{2+d}
\end{equation}
where $ f_{2k} = q^\gamma(\tau_k-t_0)$ will be denoted as $q_k^\gamma$. On averaging, and taking into account the fact that the random variables $\xi_{2k}$, $\tau_k$  and $\delta\tau_k$   are decorrelated, we get:
\begin{equation}
\langle |X_k|^{2+d} \rangle = \varepsilon_2^{2+d} \, \langle | \xi_{2k}|^{2+d} \rangle \, \langle |n_{2k} \, q_k^\gamma  \, u_{0k} |^{2+d} \rangle \,  \langle |\delta\tau_k |^{2+d}  \rangle.
\end{equation}
The time increments $\delta\tau_k$ having an exponential distribution with mean $\delta\tau$ we have $ \langle |\delta\tau_k |^{2+d}\rangle =  \delta\tau^{2+d} \Gamma(d+3)$. We assume that the moment $ \langle | \xi_{2k}|^{2+d} \rangle$ is equal to some finite value  $ \langle | \xi |^{2+d} \rangle$. Also, we assume that, for all $k$, $ \langle |n_{2k} \, q_k^\gamma  \, u_{0k} |^{2+d} \rangle$ is bounded by some positive constant $m$ independent of $N$.
This assumption is satisfied if $\gamma > 0$, since $|q_k^\gamma|$ is bounded, but might not systematically be true when $\gamma < 0$ since $|q_k^\gamma|$ diverges when $q_k$ is close to 0. However, probabilities in the case $\gamma < 0$ will not be considered in the present work.
Under these assumptions we then have:
\begin{equation}
0 < \frac{1}{S_N^{2+d}}\sum_{k=0}^{N-1} \langle| X_k|^{2+d} \rangle < N \times  \frac{\varepsilon_2^{2+d} \, \langle | \xi|^{2+d} \rangle m}{( K_1 \sqrt {\delta \tau})^{2+d}} \, \delta\tau^{2+d} \Gamma(d+3) = constant / N^{d/2}
\end{equation}
since  $\delta \tau = constant/N$. We therefore conclude that $\sum_{k} \langle| X_k|^{2+d} \rangle /S_N^{2+d} \to 0$ for any $d>0$, so that the Lyapunov condition is satisfied.

\begin{figure}
 \centerline{\includegraphics[scale=0.60]{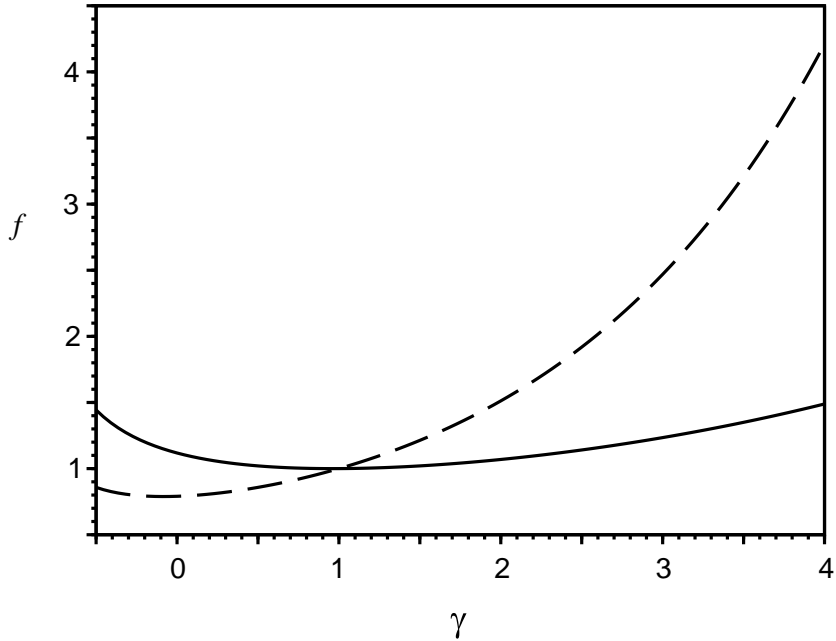}}
  \caption{Plot of $\sigma_\gamma/\sigma_1$, that is $f(\gamma)$, as a function of the exponent of the noise amplitude $\gamma$, for $\alpha/\beta=1$ and $\alpha/\beta=2$. Obtained from Eq.\ (\ref{f2gamma}).}
  \label{f_gamma}
  \end{figure}

\begin{figure}
 \centerline{\includegraphics[scale=0.60]{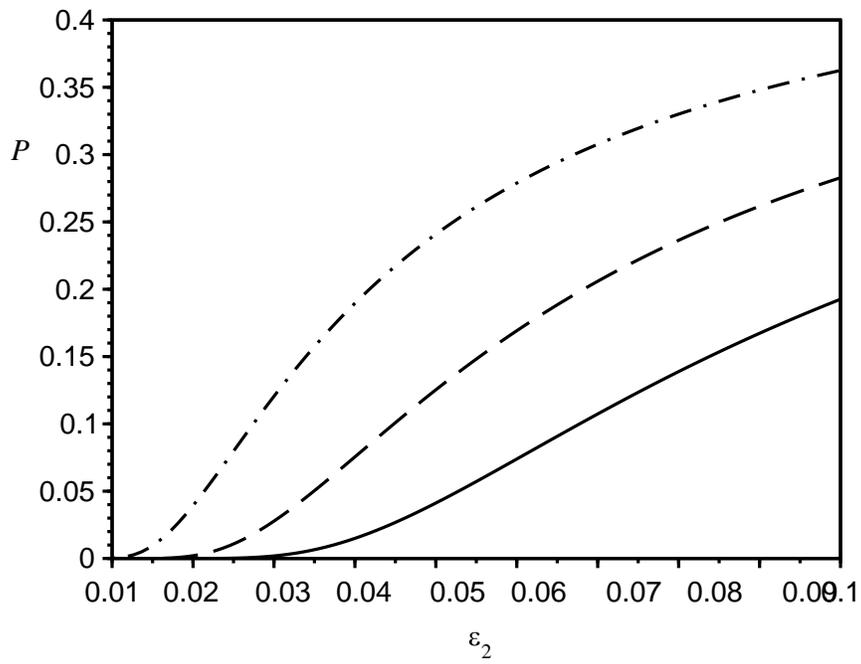}}
  \caption{Effect of the non-linearity of the amplitude of the random force $f_2(\mathbf r) = q^\gamma$, for the Duffing equation. Solid line: $\gamma=1$ (linear noise amplitude) ; dashed line:  $\gamma=2$ ; dot-dashed line:  $\gamma=3$. The friction coefficient is $\Lambda=0.008$, and $\alpha/\beta=2$.  }
  \label{P_eps_gamma1-2-3_alp2_bet1}
  \end{figure}

We will then make use of the general expression for the probability (\ref{pb1}). The standard deviation of the jump now depends on $\gamma$ and will be denoted by $\sigma_\gamma$ in the following. By making use of expression (\ref{sig2}) we get, after some algebra:
\begin{equation}
\sigma_\gamma = \sigma_1 \, f(\gamma)
\label{sig2gamma}
\end{equation}
where $\sigma_1$ is given by Eq.\ (\ref{sigDuffing}) and
\begin{equation}
 f^2(\gamma) =
\frac{15\sqrt \pi}{8} \, \frac{\Gamma(\gamma+1)}{\Gamma(\gamma+5/2)} \left( \frac{2 \alpha}{\beta} \right)^{\gamma-1}
\label{f2gamma}
\end{equation}
provided $\gamma > -1$, where $\Gamma$ denotes Euler's Gamma function.
Using this expression, one can check that the standard deviation of the jump can  be very sensitive to the non-linearity of the noise amplitude, especially if $\gamma$ is larger than a few units. As a consequence, the probability of noise-induced crossing is affected by this non-linearity.  Figure  \ref{f_gamma} shows $\sigma_\gamma/\sigma_1$, that is $f(\gamma)$, as a function of $\gamma$, for $\alpha/\beta=1$ and $\alpha/\beta=2$. We observe that the  non-linearity of the noise amplitude can significantly affect the standard deviation, especially if $\alpha/\beta$ is larger than unity. To illustrate this effect, we have calculated the probability, obtained by injecting (\ref{sig2gamma}) into  (\ref{pb1}), and using (\ref{DeltaH0Duffing}), for $\alpha/\beta=2$ (Fig.\ \ref{P_eps_gamma1-2-3_alp2_bet1}). The friction coefficient has been set to $\Lambda=0.008$. We clearly observe that the probability of noise-induced crossing is highly sensitive to $\gamma$ for these parameters. Note that, according to the values of $\alpha$ and $\beta$, one could also find cases where the probability is lowered by increasing $\gamma$.  

\newpage
 
\section{Discussion and conclusion}
\label{conclus}

We have studied how multiplicative colored noise can affect the dynamics of a system near homoclinic or heteroclinic trajectories (separatrices). By choosing a Kubo-Anderson  noise, we could perform an analytical treatment to calculate the jump of Hamiltonian for any system released near a hyperbolic saddle point of the separatrix. This jump is a Gaussian random variable, the average and variance of which have been obtained in terms of the parameters of the original system. The probability of noise-induced crossing could then be derived, and was shown to agree with numerical simulations in the classical example of the Duffing equation.

The theoretical results presented here can be applied to a wide class of dynamical systems and multiplicative noises. We have investigated the case of  multiplicative noises of the form $q^\gamma \, \xi$, i.e. with either a linear ($\gamma=1$) or a non-linear  ($\gamma \not=1$) amplitude. The latter case is of major importance for multiplicative noises, since this non-linearity can lead to very large inhomogeneities in the phase space (i.e. noise-free zones coexisting with high noise zones). This is  in high contrast with additive noises. We have illustrated this effect in the case of the Duffing oscillator and shown how the probability of noise-induced crossing could be sensitive to the exponent $\gamma$.

This work could be applied to mechanical or electrical oscillators, where many situations involve multiplicative noises, but also to transport phenomena. For example, turbulent transport of particles in fluids is often modelled by means of a stochastic term,  the intensity of which is proportional to the {\it local} turbulent characteristics at the position of the particle \cite{Kallio1989,Graham1996}. This random term is therefore multiplicative and non-linear in terms of the state variables of the system (here the coordinates of the particle). Zones with high turbulence intensity often coexist with laminar zones, and the turbulent dispersion of particles is therefore strongly affected by these inhomogeneities. In the case of plane flows, the theory developed here could be readily used to study noise-induced separatrix crossing. In particular, it can give  the proportion of particles entering a recirculation cell under the effect of turbulent dispersion. Other advection problems containing a deterministic and a random component of this kind, like the transport of micro-organisms in the presence of chemical signals, can be studied by means of the theory presented in this paper. They constitute the next steps of this project.
 
Finally, it is well known that non-linear friction also can significantly affect the dynamics of Duffing oscillators  \cite{Ravindra1994}\cite{Sharma2012}. This effect, which was out of the scope of the present study, can be readily accounted for through the deterministic part of the jump $\Delta H^0$. The combined effects of both non-linear  noise and non-linear friction on the probability of  noise-induced crossing  can then be obtained analytically. A detailed analysis of such systems will be performed in the near future.


  \end{document}